\documentclass[pra,twocolumn,aps,showpacs,superscriptaddress]{revtex4-1}
\usepackage{epsfig}
\usepackage{subfigure}
\usepackage{bm}
\usepackage{amsmath}
\usepackage{hyperref}
\begin{document}
\title{Bose-Fermi mixtures in the molecular limit}
\author{Andrea Guidini}
\affiliation{School of Science and Technology, Physics Division, 
University of Camerino, Via Madonna delle Carceri 9, I-62032 Camerino, Italy}
\author{Gianluca Bertaina}
\affiliation{Dipartimento di Fisica, Universit\`a degli Studi di Milano, via Celoria 16, I-20133 Milano, Italy}
\author{Elisa Fratini}
\affiliation{The Abdus Salam International Centre for Theoretical Physics, 34151 Trieste, Italy}
\author{Pierbiagio Pieri}
\affiliation{School of Science and Technology, Physics Division, 
University of Camerino, Via Madonna delle Carceri 9, I-62032 Camerino, Italy}
\date{\today}

\begin{abstract}
We consider  a Bose-Fermi mixture in the molecular limit of the attractive interaction between fermions and bosons.  For a boson density smaller or equal to the fermion density, we show analytically how a T-matrix approach for the constituent bosons and  fermions  recovers the expected physical limit of a Fermi-Fermi mixture of molecules and atoms. In this limit, we derive simple expressions for the self-energies, the momentum distribution function, and the chemical potentials.
By extending these equations to a trapped system,  we determine how to tailor the experimental parameters of a Bose-Fermi mixture in order to enhance the  {\em  indirect Pauli exclusion effect} on the boson momentum distribution function. For the homogeneous system, we present finally a Diffusion Monte Carlo simulation which confirms the occurrence of such a peculiar effect.
\end{abstract}

\pacs{03.75.Ss,03.75.Hh,32.30.Bv,74.20.-z}
\maketitle

\section{Introduction}\label{intro}

Bose-Fermi mixtures with a tunable boson-fermion attraction have been object of active theoretical
 \cite{Pow05,Dic05,Sto05,Avd06,Pol06,Roethel07,Bar08,Pol08,Bor08,Mar08,Wat08,Fra10,Yu11,Lud11,Song11,Fra12,Yam12,And12,Ber13,Fra13,Sog13}
 and experimental~\cite{Osp06,Osp06b,Zir08,Ni08,Wu11,Wu12,Park12,Heo12,Cum13,Bloom13}  investigation over the last few years.
 
 Previous theoretical studies of these systems have shown that for a sufficiently strong attraction between fermions and bosons, the boson condensation is completely 
 suppressed in mixtures where the boson density  $n_{\rm B}$ is smaller or equal to the fermion density $n_{\rm F}$. This complete suppression of condensation occurs even at zero temperature, and is associated  to pairing of bosons with fermions into composite fermions. Since the binding occurs in a medium, the paired state formed by one boson and one fermion is influenced by the presence of the remaining particles, and its composite nature can manifest in appropriate thermodynamic or dynamic quantities.   
Clearly, when the attraction is increased further, the internal degrees of freedom of the composite fermions  are progressively  frozen and the original  Bose-Fermi mixture becomes effectively a Fermi-Fermi mixture of molecules and atomic (unpaired) fermions.
 This kind of evolution has been studied already by us with a T-matrix diagrammatic formalism~\cite{Fra10,Fra12,Fra13} and with Fixed-node Diffusion Monte Carlo~\cite{Ber13}.

Aim of the present paper is to show {\em analytically} how  a T-matrix diagrammatic approach, which is formulated in terms of the constituent bosons and fermions, reconstructs the appropriate description in terms of molecules and unpaired fermions when the attraction is sufficiently large. In this limit, we will derive simple expressions for the bosonic and fermionic self-energies, momentum distribution functions and chemical potentials. Special attention will be devoted to the momentum distribution functions. Indeed, one very interesting feature found previously by us in a Bose-Fermi mixture is the presence, under appropriate conditions, of a region at low momenta with zero occupancy in the bosonic momentum distribution function. The presence of this region was interpreted as an indirect effect on the bosonic distribution of the  Pauli exclusion principle acting on the unpaired and composite fermions.  The analytic expression that we will derive in this paper for the momentum distribution function will make such {\em indirect Pauli exclusion effect}  on the bosonic component particularly transparent. 

The use of these simple equations will allow us to incorporate easily also the effect of an external  trapping potential. We will calculate then the density profiles and the momentum distribution functions for the trapped system.  We will focus in particular in determining the ideal experimental parameters that maximize the indirect Pauli exclusion effect, as to make it  possibly observable in future experiments with Bose-Fermi mixtures. In this respect, we will see that mixtures where the bosons are light compared to the fermions are particularly promising. 

The paper is organized as follows.
 In section \ref{formalism} we derive the asymptotic expressions for the pair propagator, self-energies, momentum distributions functions, and chemical potentials, that are obtained in the molecular limit of the Bose-Fermi attraction by starting from the T-matrix self-energies. A comparison between the asymptotic expressions and the corresponding T-matrix results is reported in Sec.~\ref{resultshomo}. In Sec.~\ref{qmc} we present Quantum Monte Carlo estimates for the bosonic momentum distribution function and compare them to the T-matrix results and asymptotic expressions. In Sec.~\ref{trapped} we include the effect of a trapping potential in the asymptotic expressions derived in Sec.~\ref{formalism}, and discuss the visibility of the indirect Pauli exclusion effect in Bose-Fermi mixtures of current experimental relevance.
 Section~\ref{conclusions} presents finally our concluding remarks.
 
\section{Derivation of the asymptotic equations in the molecular limit}\label{formalism}
\subsection{Preliminaries}
We consider a mixture of single-component fermions and bosons, with the boson-fermion interaction described by a contact interaction, as it can be realized with a (broad) Fano-Feshbach resonance tuning the boson-fermion scattering length $a$ of an ultracold Bose-Fermi mixture.
We will be interested in particular  in the molecular limit of this system, namely the limit where the binding energy $\epsilon_0$  of the two-body boson-fermion bound state is the dominant energy scale. For the contact potential $\epsilon_0=1/(2 m_r a^2)$, where $m_r=m_{\rm B} m_{\rm F}/(m_{\rm B}+m_{\rm F})$  is the reduced mass determined by the boson and fermion masses $m_{\rm B}$ and $m_{\rm F}$, and we have set $\hbar=1$.
 The repulsive potential between bosons, which is necessary for the stability of the system in the resonance region, can be dropped out from our consideration in the molecular limit of interest to the present paper.

A natural length scale of our system, where fermions are the majority species, is provided by the inverse of the Fermi wave-vector $k_{\rm F}\equiv (6 \pi^2 n_{\rm F})^{1/3}$   ($n_{\rm F}$ being the fermion number density).  
One may use then the dimensionless coupling parameter $g=(k_{\rm F} a)^{-1}$ to describe the strength of the interaction.  
In terms of this parameter, the molecular limit corresponds to the condition $g \gg 1$, such that the radius of the bound state (which coincides with the scattering length $a$, for $a$  positive) is much smaller than the average interparticle distance ($\propto k_{\rm F}^{-1}$). [Note that in some of our previous works \cite{Fra10,Fra12,Fra13} we used a different definition of $k_{\rm F}$ (in terms of the total density $n=n_{\rm F}+ n_{\rm B}$, $n_{\rm B}$ being the boson number density), which coincides with the present one only for  $n_{\rm F}=n_{\rm B}$.]

 The thermodynamic and spectral properties of a Bose-Fermi mixture in the normal phase (i.e. above the condensation critical temperature) were studied in  our previous works~\cite{Fra10,Fra12,Fra13} by using a T-matrix approximation for the self-energies.  The corresponding equations for the bosonic and fermionic self-energies $\Sigma_{\rm B}$ and $\Sigma_{\rm F}$ at finite temperature read (setting the Boltzmann constant $k_{\rm B}=1$):
\begin{eqnarray}\label{selfb}
\Sigma_{\rm B}({\bf k},\omega_{\nu})&=&-\int\!\!\frac{d {\bf P}}{(2\pi)^{3}} T \sum_{m}\Gamma({\bf P},\Omega_{m})\nonumber\\
&\times& G_{\rm F}^{0}({\bf P}-{\bf k},\Omega_{m}-\omega_{\nu})\\
\label{selff}
\Sigma_{\rm F}({\bf k},\omega_{n})&=&\int\!\!\frac{d {\bf P}}{(2\pi)^{3}} T \sum_{\rm m} \Gamma({\bf P},\Omega_{m})\nonumber\\
&\times&G_{\rm B}^{0}({\bf P}-{\bf k},\Omega_{m}-\omega_{n})
\end{eqnarray}
where the pair propagator  $\Gamma({\bf P},\Omega_{m})$ is given by
\begin{eqnarray}
&&\Gamma({\bf P}, \Omega_{m})=- \left\{\frac{m_{r}}{2\pi a}+\int\!\!\frac{d{\bf p}}{(2\pi)^{3}}
\right.\nonumber\\
&&\times \left.\left[\frac{1-f(\xi^{\rm F}_{{\bf P}-{\bf p}})+b(\xi^{\rm B}_{{\bf p}})}{\xi^{\rm F}_{{\bf P}-{\bf p}}+\xi^{\rm B}_{\bf p}-i\Omega_{m}}
-\frac{2m_{r}}{p^{2}} \right]\right\}^{-1}.
\label{gamma}
\end{eqnarray}

In the above expressions, $\omega_{\nu}=2\pi\nu T$  and $\omega_{n}=(2n+1)\pi T$, $\Omega_{m}=(2m+1)\pi T$ 
are bosonic and fermionic Matsubara frequencies, respectively, ($\nu,n,m$ being integer numbers), while $f(x)$ and $b(x)$ are the Fermi and Bose distribution functions at temperature $T$  [$f(x)=(e^{x/T}+1)^{-1}$, $b(x)=(e^{x/T}-1)^{-1}$].   In Eq.~(\ref{gamma})  $\xi^s_{\bf p}=p^2/2m_s-\mu_s$ is the free dispersion relative to the chemical potential $\mu_s$ for the species $s={\rm B},{\rm F}$, while the  bare Green's functions appearing in Eqs.~(\ref{selfb}) and (\ref{selff}) are given by $G^{0}_{\rm B}({\bf k},\omega_{\nu})^{-1}=i\omega_{\nu}-\xi^{\rm B}_{{\bf k}}$ and $G^{0}_{\rm F}(\mathbf{k},\omega_{n})^{-1}=i\omega_{n}-\xi^{\rm F}_{\mathbf{k}}$.

The self-energies (\ref{selfb}) and (\ref{selff}) determine the dressed Green's functions $G_{s}$ via the Dyson's equation $G_{s}^{-1}=G_{s}^{0 \; -1}-\Sigma_{s}$. The dressed Green's functions $G_s$, in turn,  allow one to calculate the boson and fermion momentum distribution functions $n_{s}({\bf k})$ through the equations:
\begin{eqnarray}\label{nbq}
n_{\rm B}({\bf k})&=&- T \sum_{\nu}G_{\rm B}({\bf k},\omega_{\nu})\,e^{i\omega_{\nu} 0^+}\\
\label{nfk}
n_{\rm F}({\bf k})&=& T \sum_{n}G_{\rm F}({\bf k},\omega_{n})\,e^{i\omega_{n} 0^+}\,,
\end{eqnarray}
from which the boson and fermion number densities are obtained by integrating over momenta.

The full numerical solution of Eqs.~(\ref{selfb})$-$(\ref{nfk}) was tackled in our previous works  \cite{Fra10,Fra12}. Here we are interested in deriving analytic expressions 
in the molecular limit of the interaction $g$. In this limit the binding energy $\epsilon_0$ is the largest energy scale: $\epsilon_0 \gg T, E_{\rm F}$, with $E_{\rm F}\equiv k_{\rm F}^2/(2m_{\rm F})$.  
In addition, for the mixtures with $n_{\rm B} \le n_{\rm F}$ of interest to the present paper, the bosonic chemical potential $\mu_{\rm B}$ approaches $-\epsilon_0$ in the molecular limit, 
and is thus large and negative, while the fermion chemical potential remains of  the order of the Fermi energy, and is then small compared to the binding energy.  This hierarchy between different energy scales will allow us to derive the asymptotic expressions in the molecular limit.  
  
\subsection{The pair propagator}
We focus first on the pair propagator $\Gamma(\mathbf{P},\Omega_{m})$. In order to perform the frequency sum in  Eqs.~(\ref{selfb}) and (\ref{selff}) for the self-energies, we
need to know the analytic properties of the extension $\Gamma(\mathbf{P},z)$ of the pair propagator to the  whole complex (frequency) plane. The analytic extension 
$\Gamma(\mathbf{P},z)$ is defined by replacing $i \Omega_{m} \to z$ on the right-hand side of Eq.~(\ref{gamma}). 
It is easy to verify directly from Eq.~(\ref{gamma}) that  $\Gamma(\mathbf{P},z)$ has a branch-cut on the real axis for ${\rm Re}\, z \ge -2\mu + P^2/(2M)$, where  
$M\equiv m_{\rm B}+m_{\rm F}$~\cite{footnote1}. In addition, for sufficiently strong attraction, the pair propagator $\Gamma(\mathbf{P},z)$ has a pole, which is associated to molecular binding.
In order to determine this pole, we first integrate the terms in Eq.~(\ref{gamma}) that do not contain the Fermi or Bose functions [this can be done for ${\rm Im} z \neq 0$ or
 for ${\rm Im} z = 0$ and 
${\rm Re}\,  z\, \le -2\mu + P^2/(2M)$].   
 The  pair propagator can be written then
\begin{eqnarray}
\label{gamma2}
\Gamma(\mathbf{P},z)&=&- \left\{\frac{m_{r}}{2\pi a}-\frac{m_{r}^{3/2}}{\sqrt{2}\,\pi}\,\sqrt{\frac{P^2}{2M}-2\mu-z}\nonumber \right. \\ 
&+&\left. I_{\rm B}(\mathbf{P},z)-I_{\rm F}(\mathbf{P},z)\phantom{\frac{1}{2}}\!\!\!\!\!\right\}^{-1},
\end{eqnarray}
where $\mu\equiv (\mu_{\rm B}+\mu_{\rm F})/2$,  while 
$I_{\rm F}(\mathbf{P},z)$ and $I_{\rm B}(\mathbf{P},z)$ are defined by:
\begin{eqnarray}\label{ib}
I_{\rm B}(\mathbf{P},z)&\equiv & \int\frac{d\mathbf{p}}{(2\pi)^{3}}\,\frac{b(\xi^{\rm B}_{\mathbf{p}})}{\xi^{\rm F}_{\mathbf{P}-\mathbf{p}}+\xi^{\rm B}_{\mathbf{p}}-z},\\
\label{if}
I_{\rm F}(\mathbf{P},z)&\equiv & \int\frac{d\mathbf{p}}{(2\pi)^{3}}\,\frac{f(\xi^{\rm F}_{\mathbf{P}-\mathbf{p}})}{\xi^{\rm F}_{\mathbf{P}-\mathbf{p}}+\xi^{\rm B}_{\mathbf{p}}-z}.
\end{eqnarray}

The term $I_{\rm B}(\mathbf{P},z)$ is suppressed  exponentially by the Bose function  $\propto \exp(-\epsilon_0/T)$  and  thus does not contribute to the pair propagator in the molecular limit. The term $I_{\rm F}({\bf P},z)$ can instead be expanded in powers of $\epsilon_0^{-1}$, assuming $P^2/(2M), |z| \ll \epsilon_0$ .
The leading term is given by:
\begin{equation}\label{appequ2}
I_{\rm F}^0=\int\!\!{\frac{d \bf p}{(2\pi)^3}}\frac{f(\xi ^{\rm F}_{\mathbf{P}})}{\epsilon_ 0}\equiv\frac{n^0_{\mu_ F}}{\epsilon _0},
\end{equation}
while inclusion of the next-to-leading term yields
\begin{equation}\label{appequ6}
I_{\rm F}({\bf P},z)=I_{\rm F}^0-\delta I_{\rm F}^0+\frac{I_{\rm F}^0}{\epsilon_ 0}\left(z +\mu _{\rm CF}-\frac{P^2}{2m_{\rm B}}\right),
\end{equation}
where we have introduced the composite-fermion chemical potential $\mu_{\rm CF}=2\mu+\epsilon_0$, while 
\begin{equation}\label{appequ4}
\delta I_{\rm F}^0=\frac{1}{2m_r \epsilon _0^2}\int\!\! 
\frac{{d \bf p}}{(2\pi)^3}
 f(\xi ^{\rm F}_{\bf p})p^2 .
\end{equation}
At $T=0$, $n^0_{\mu_ {\rm F}}= \frac{k_{\mu_ {\rm F}}^3}{6\pi^2}$ with $k_{\mu_{\rm F}}=\sqrt{2 m_{\rm F} \mu_{\rm F}}$, while  $\delta I_{\rm F}^0=3 m_{\rm F}   \mu _{\rm F} I_{\rm F}^0/(5 m_r \epsilon _0)$.

The pole of  $\Gamma$ is then determined by the equation:  
\begin{equation}\label{appequ10}
z-\frac{P^2}{2M}+2\mu+\epsilon_ 0\left[1-\frac{2\pi a}{m_r}I_{\rm F}(\mathbf{P},z)\right]^2=0,
\end{equation}
which, by using the expansion (\ref{appequ6}) and neglecting terms of order $\epsilon_0^{-2}$ at least, yields 
\begin{equation}\label{appequ13}
z=\frac{P^2}{2M^*}-\mu _{\rm CF}+\Sigma_{\rm CF},
\end{equation}
where the molecule effective mass $M^*$  is given by 
\begin{eqnarray}
M^*&=&M \left(1+I_{\rm F}^0\frac{4\pi a}{m_r}\frac{m_{\rm F}}{m_{\rm B}}\right)\\
&=&M \left[1+\frac{4}{3\pi}\frac{m_{\rm F}}{m_{\rm B}}(k_{\mu_{\rm F}} a)^3\right]\;\;\;\; (T=0) \label{equ14},
\end{eqnarray}
while the self-energy correction $\Sigma_{\rm CF}$ is 
\begin{eqnarray}
\Sigma_{\rm CF}&=&\frac{4\pi a}{m_r}\epsilon_ 0 (I_{\rm F}^0-\delta I_{\rm F}^0)\\
&=& \frac{4\pi a}{m_r} n^0_{\mu_{\rm F}}\left[1-\frac{3}{5}(k_{\mu_{\rm F}} a)^2\right] \;\;\;\; (T=0) .
\end{eqnarray}
The Eq.~(\ref{appequ13}) for the pole determines then the dressed dispersion of the composite-fermion 
\begin{equation}
\label{pcf}
\tilde{\xi}^{\rm CF}_{\bf P}=\frac{P^2}{2M^*}-\mu _{\rm CF}+\Sigma_{\rm CF},
\end{equation}
and the associated composite-fermion Fermi momentum $P_{\rm CF}$, defined by the equation $\tilde{\xi}^{\rm CF}_{P_{\rm CF}}=0$.
Note that the leading order term of the self-energy  $\Sigma_{\rm CF}$ takes into account the interaction between the molecules and the unpaired fermions (with approximate density $n^0_{\mu_{\rm F}}$) with the Born approximation value for the molecule-fermion scattering length $a_{\rm DF}$:
\begin{equation} 
a_{\rm DF}=\frac{(1 + m_{\rm F}/m_{\rm B})^2}{1+ m_{\rm F}/2m_{\rm B}} a .
\end{equation}
The subleading correction to $\Sigma_{\rm CF}$ as well as the correction (\ref{equ14}) to the bare mass of the molecules are instead due to the composite nature
of the molecules.

Finally, the residue $w({\bf P})$  at the pole of $\Gamma$ is given by
\begin{eqnarray}
w({\bf P})&=& \lim_{z\to\tilde{\xi}^{\rm CF}_{\bf P}} (z-\tilde{\xi}^{\rm CF}_{\bf P})\Gamma(z,{\bf P})\\
&=&-\frac{2\pi}{am_r^2}\frac{1-\frac{2\pi a }{m_r}I_{\rm F}(\mathbf{P},\tilde{\xi}^{\rm CF}_{\bf P})}{1-\frac{4\pi a}{m_r}I_{\rm F}^0}\\
&\simeq &\label{residue}
-\frac{2\pi}{am_r^2}\left(1+\frac{2\pi a}{m_r}I_{\rm F}^0\right)\equiv -w_0,
\end{eqnarray}
where  in the last line we have neglected again terms of order $\epsilon_0^{-2}$. We see in Eq.~(\ref{residue}) that  in the molecular limit the dependence on ${\bf P}$ of the residue is negligible even at next-to-leading order.

\subsection{Bosonic self-energy and momentum distribution function}\label{bosonic}
The sum over the fermionic frequency $\Omega_{m}$ in Eq.~(\ref{selfb}) for the bosonic self-energy can be performed by transforming it in a contour integration in the complex $z$ plane, as usually done when summing over Matsubara frequencies (see, e.g., chap.~7 of Ref.~\cite{fetter}).
One obtains three contributions associated to the different singularities of $\Gamma$  and $G_{\rm F} ^0$ in the complex plane: the simple poles of $\Gamma$ and $G_{\rm F} ^0$ and the integral along the branch-cut of $\Gamma({\bf P},z)$.  This integral is, however, suppressed exponentially by the Fermi function $f(z)$ which appears when transforming the discrete sum in a contour integration.
 Indeed, we have seen above that the branch-cut is present for  ${\rm Re}\, z \ge -2\mu + P^2/(2M)$.  Since $2\mu \simeq -\epsilon_0$ in the molecular limit, it follows immediately that the integral along the cut is suppressed exponentially at finite temperature (and is vanishing at $T=0$). 

 The contributions from the poles of $\Gamma$ and $G_{\rm F} ^0$ yield then
\begin{equation}
\label{equ39a}
\Sigma_{\rm B}({\bf k},\omega_{\nu})=w_0 \int\!\!\!\frac{d{\bf P}}{(2\pi)^3}\,\frac{f(\tilde{\xi}^{\rm CF}_{\bf P})-f(\xi^F_{{\bf P}-{\bf k}})}{\tilde{\xi}^{\rm CF}_{\bf P}- \xi^F_{{\bf P}-{\bf k}} 
-i \omega_{\nu}} .
\end{equation}

Note how in the molecular limit the boson self-energy (\ref{equ39a}) acquires the form determined  by the virtual recombination of the boson with a fermion to form  a molecule, with probability amplitude $\sqrt{w_0}$, followed by the decay of the virtual molecule into its constituent fermions and bosons (with the same probability amplitude).

We pass now to the calculation of the boson momentum distribution function, as determined by Eq.~(\ref{nbq}). We first notice that  in the molecular limit we are allowed to expand perturbatively the Dyson's equation $G_{\rm B}({\bf k},\omega_{\nu})=[G_{\rm B}^0({\bf k},\omega_{\nu})^{-1}-\Sigma_{\rm B}({\bf k},\omega_{\nu})]^{-1}$ since  $\mu _{\rm B}$, and therefore the  relevant range of values of  $\omega_{\nu}$  and ${\bf k}^2/(2 m_{\rm B})$ inside the free boson propagator,  are of order $\epsilon_0$, while $\Sigma_{\rm B}$ is of order $\epsilon_0^{1/2}$ (because of the residue $w_0$, which is of order  $\epsilon_0^{1/2}$ ).

The expansion of the Dyson's equation to first order then yields:
\begin{equation}\label{appequ21}
G_{\rm B}({\bf k},\omega_{\nu}) \simeq G_{\rm B}^0({\bf k},\omega_{\nu})+G_{\rm B}^{0}({\bf k},\omega_{\nu})^2\Sigma _{\rm B} ({\bf k},\omega_{\nu}).
\end{equation}
The first term on the right-hand-side of the above equation yields again an exponentially small contribution when summed over  $\omega_{\nu} $. 
 By inserting the expression (\ref{equ39a}) for the self-energy in Eq.~(\ref{appequ21}) and summing over $\omega_{\nu}$ one gets then:
\begin{eqnarray}\label{appequ24}
n_{\rm B}(k) &=& w_0\! \int\!\!\frac{d\bf P}{(2\pi)^3}\frac{b(\tilde{\xi}^{\rm CF} _{\bf P}-\xi ^{\rm F}_{\mathbf{P}-{\bf k}})[f(\tilde{\xi}^{\rm CF} _{\mathbf{P}})-f(\xi ^{\rm F}_{\mathbf{P}-{\bf k}})]}
{(\xi ^{\rm B}_{{\bf k}}+\xi ^{\rm F}_{\mathbf{P}-{\bf k}}-\tilde{\xi}^{ \rm CF} _{\mathbf{P}})^2}\nonumber\\
&=& w_0\! \int\!\!\frac{d\bf P}{(2\pi)^3}
\frac{f(-\xi ^{\rm F}_{\mathbf{P}-{\bf k}}) f(\tilde{\xi}^{\rm CF} _{\bf P})}
{(\xi ^{\rm B}_{\bf k}+\xi ^{\rm F}_{{\bf P}-{\bf k}}-\tilde{\xi}^{\rm CF} _{\bf P})^2}\\
&=& w_0\! \int\!\!\frac{d\bf P}{(2\pi)^3}
\frac{\Theta(\xi ^{\rm F}_{\mathbf{P}-{\bf k}}) \Theta(P_{\rm CF}-P)}
{(\xi ^{\rm B}_{\bf k}+\xi ^{\rm F}_{{\bf P}-{\bf k}}-\tilde{\xi}^{\rm CF} _{\bf P})^2}\;\;\;\;(T=0).\nonumber\label{bb}\\
&&
\end{eqnarray}
The expressions (\ref{appequ24}) and (\ref{bb}) show clearly the effect of the Fermi statistics obeyed by the molecules and unpaired fermions on the bosonic momentum distribution function. In particular, at $T=0$ the two $\Theta$ functions in Eq.~(\ref{bb}) require simultaneously $P<P_{\rm CF}$ and $|{\bf P}-{\bf k}| > k_{\mu_{\rm F}}$.
 As a result, when  $k_{\mu_{\rm F}} > P_{\rm CF}$,  $n_{\rm B}(k)=0$ for $k < k_{\mu_{\rm F}}-P_{\rm CF}$. We see therefore that, for sufficiently low boson concentration, such that $k_{\mu_{\rm F}} > P_{\rm CF}$, the formation of the molecules depletes completely the bosonic momentum distribution at low momenta. In particular, by using 
 the asymptotic expressions for the chemical potentials derived below, one can see that, to leading order in the molecular limit, $k_{\mu_{\rm F}}$ corresponds to the radius of the  Fermi sphere of the unpaired fermions, with density $n_{\rm F}-n_{\rm B}$, while $P_{\rm CF}$ corresponds to the radius of the Fermi sphere of the composite fermions, with density $n_{\rm B}$. It then follows  that  the condition $n_{\rm F}-n_{\rm B} > n_{\rm B}$ must be fulfilled in order to have  $k_{\mu_{\rm F}} > P_{\rm CF}$, and therefore the presence of the empty region at low momenta.
We note further that a partial suppression of the bosonic momentum distribution at low momenta was found also for  weakly-interacting Bose-Fermi mixtures in the perturbative analysis of Ref.~\onlinecite{Viv02}. We see that in the opposite (molecular) limit this effect is made extreme, yielding for $n_{\rm B} < n_{\rm F}/2$ to a complete suppression of the occupancy at low momenta. 

\subsection{Fermionic self-energy and momentum distribution function}\label{fermionic}
The calculation of the fermionic self-energy from Eq.~(\ref{selff}) proceeds along the same lines as for the bosonic self-energy, with the only difference that in this case there is just the pole of  $\Gamma$ to be considered, since the pole of  $G_{\rm B} ^0$ is suppressed exponentially.  
One obtains then
\begin{eqnarray}\label{appequ49}
\Sigma _{\rm F} (\mathbf{k},\omega _n)&=&-w_0\int\!\!\!\frac{d\bf P}{(2\pi)^3}\frac{f(\tilde{\xi} ^{\rm CF} _{\mathbf{P}})}{\tilde{\xi} ^{\rm CF} _{\mathbf{P}}-\xi ^{\rm B}_{\mathbf{P}-\mathbf{k}} -i\omega _n}.
\end{eqnarray}
The presence of $\mu_{\rm B}$ in the denominator of the expression (\ref{appequ49}) for the  fermionic self-energy makes it to behave in the molecular limit like $w_0\epsilon_0^{-1}\sim\epsilon_0^{-1/2}$.  We are allowed then to expand the Dyson's equation also for the fermionic Green's function. Before doing this, it is useful to introduce a procedure which accelerates the convergence of the expansion in the fermionic case. Indeed, in this case,  there is a range of $k$ close to the Fermi step, and of frequencies close to zero such that $G_{\rm F}^0(\mathbf{k},\omega_{n})^{-1}$ may be comparable or even smaller than $\Sigma _{\rm F}(\mathbf{k},\omega_{n})$, thus invalidating the expansion of the Dyson's equation in this region.  (In the bosonic case, for which the boson chemical potential is negative and large, the self-energy is instead always much smaller than $G_{\rm B}^{0}\, ^{-1}$.)    Before expanding, we thus add and subtract in the denominator of the Dyson's equation the quantity $\Sigma_{\rm F} ^0 \equiv {\rm Re} \Sigma_{\rm F}^{\rm R} (k_{\rm UF},\omega=0)$, where $k_{\rm UF}$ corresponds to the position of the Fermi step  of $G_{\rm F}$ as defined by the equation 
$$k^2/(2m_{\rm F})-\mu _{\rm F} + {\rm Re} \Sigma _{\rm F}^{\rm R} (k, \omega =0)=0,$$ and $\Sigma_{\rm F}^{\rm R}(\mathbf{k}, \omega)$ is the analytic continuation of the self-energy to the real axis (obtained with the replacement $i\omega_{n} \to \omega + i 0^+$). 
In practice, in the molecular limit of our interest, $k_{\rm UF}=[6\pi^2(n_{\rm F}-n_{\rm B})]^{1/3}$, as we will see below.
We have then:
\begin{eqnarray}
\label{dysonfermi}
G_{\rm F}(\mathbf{k},\omega_{n})&=&\frac{1}{\tilde{G} _{\rm F} ^0(\mathbf{k},\omega_{n})^{-1}-\tilde{\Sigma} _{\rm F} (\mathbf{k},\omega_{n})},
\end{eqnarray}
where $\tilde{G}^0_{\rm F} (\mathbf{k},\omega_{n})^{-1}=i\omega _n -\tilde{\xi}^ {\rm F}_{\bf k}$
with $\tilde{\xi} ^{\rm F}_{\mathbf{k}}= k^2/(2m_{\rm F})-\mu _{\rm F} + \Sigma_{\rm F} ^0$, while $\tilde{\Sigma} _{\rm F} (\mathbf{k},\omega_{n}) =\Sigma _{\rm F} (\mathbf{k},\omega_{n})-\Sigma _{\rm F} ^0$. 
The expansion of the Dyson's equation (\ref{dysonfermi}) improves on that of the original equation.
 Indeed, in the region where $\tilde{G}_{\rm F}  (\mathbf{k},\omega_{n}) ^{-1}$ is small or vanishing, $\tilde{\Sigma}_{\rm F} (\mathbf{k},\omega_{n})$ is also vanishing. In addition, it is easy to check from Eq.~(\ref{appequ49}) that $\tilde{\Sigma}_{\rm F} (\mathbf{k},\omega_{n})$ is of order
$\epsilon_0^{-3/2}$ (while $\Sigma _{\rm F} (\mathbf{k},\omega _n)$ is of order $\epsilon_0^{-1/2}$), thus accelerating the convergence of the expansion of the Dyson's equation.
We thus have 
\begin{equation}
G_{\rm F} (\mathbf{k},\omega_{n})\simeq \tilde{G}_{\rm F}^0(\mathbf{k},\omega_{n}) + \tilde{G}_{\rm F}^0  (\mathbf{k},\omega_{n}) ^2 \tilde{\Sigma}_{\rm F} (\mathbf{k},\omega_{n}),
\end{equation}
from which one obtains
\begin{eqnarray}
n_{\rm F}(k) = f(\tilde{\xi} ^{\rm F}_{\mathbf{k}}) +T \sum_{n}\tilde{G}_{\rm F}^{0}({\bf k},\omega _n)^2 \tilde{\Sigma}_{\rm F} (\mathbf{k},\omega_{n})\phantom{aaaaaaa}&&\\
=f(\tilde{\xi} ^{\rm F}_{\mathbf{k}}) +T \sum_{n}\tilde{G}_{\rm F}^{0}({\bf k},\omega _n)^2 \Sigma_{\rm F} (\mathbf{k},\omega_{n}) - \Sigma_0^{F} f '(\tilde{\xi} ^{\rm F}_{\mathbf{k}}).\phantom{aaa}&&
\label{aa}
\end{eqnarray}
By using Eq.~(\ref{appequ49}), one gets then:
\begin{eqnarray}
T \sum_{n}\tilde{G}_{\rm F}^{0}({\bf k},\omega _n)^2 \Sigma_{\rm F} (\mathbf{k},\omega_{n}) =-w_0\int\!\!\!\frac{d\bf P}{(2\pi)^3} f(\tilde{\xi} ^{\rm CF} _{\mathbf{P}})\phantom{aa}&&\nonumber\\
\times\left[ \frac{-f(\tilde{\xi} ^{\rm CF} _{\mathbf{P}}-\xi ^{\rm B}_{\mathbf{P}-\mathbf{k}})+f (\tilde{\xi}^{\rm F}_{\mathbf{k}})}{(\tilde{\xi}^{\rm CF} _{\mathbf{P}}-\xi ^{\rm B}_{\mathbf{P}-\mathbf{k}} -\tilde{\xi}^{\rm F}_{\mathbf{k}})^2}
+ \frac{f '(\tilde{\xi}^{\rm F}_{\mathbf{k}})}{\tilde{\xi} ^{\rm CF} _{\mathbf{P}}-\xi ^{\rm B}_{\mathbf{P}-\mathbf{k}} -\tilde{\xi}^{\rm F}_{\mathbf{k}}}  \right]\phantom{aa}&&
\end{eqnarray}
Note that, neglecting exponentially small terms  in the molecular limit, $f(\tilde{\xi} ^{\rm CF} _{\mathbf{P}}-\xi ^{\rm B}_{\mathbf{P}-\mathbf{k}})=1-f(-\tilde{\xi} ^{\rm CF} _{\mathbf{P}}+\xi ^{\rm B}_{\mathbf{P}-\mathbf{k}})
\simeq 1 $. In addition, at $T=0$, where $f '(\tilde{\xi} ^{\rm F}_{\mathbf{k}})=-\delta(\tilde{\xi} ^{\rm F}_{\mathbf{k}})$,
\begin{equation}
-w_0\int\!\!\!\frac{d\bf P}{(2\pi)^3}\frac{f(\tilde{\xi} ^{\rm CF} _{\mathbf{P}}) f'(\tilde{\xi}^{\rm F}_{\mathbf{k}})}{\tilde{\xi}^{\rm CF} _{\mathbf{P}}-\xi ^{\rm B}_{\mathbf{P}-\mathbf{k}} -\tilde{\xi}^{\rm F}_{\mathbf{k}}}= -\Sigma_0^{F} \delta(\tilde{\xi} ^{\rm F}_{\mathbf{k}}),
\end{equation}
which cancels exactly with the last term on the r.h.s of Eq.~(\ref{aa}). At finite $T$ this cancellation holds only approximately, the difference being a term of order $T/\epsilon_0^{3/2}$, which is anyway negligible in the molecular limit.

We thus obtain for the fermionic momentum distribution function in the molecular limit:
\begin{eqnarray}
\label{cc}
&&n_{\rm F}(k) = f(\tilde{\xi} ^{\rm F}_{\mathbf{k}})+ f(-\tilde{\xi}^{\rm F}_{\mathbf{k}})\!\int\!\!\!\frac{d\bf P}{(2\pi)^3} \frac{w_0 f(\tilde{\xi} ^{\rm CF} _{\mathbf{P}})}{(\tilde{\xi}^{\rm CF} _{\mathbf{P}}-\xi ^{\rm B}_{\mathbf{P}-\mathbf{k}} -\tilde{\xi}^{\rm F}_{\mathbf{k}})^2},\nonumber\\
&&
\end{eqnarray}
which at $T=0$ becomes:
\begin{eqnarray}
\label{dd}
&&n_{\rm F}(k) = \Theta(k_{\rm UF}-k) \nonumber\\
&& +  \Theta(k-k_{\rm UF})\!\int\!\!\!\frac{d\bf P}{(2\pi)^3} \frac{w_0 \Theta(P_{\rm CF}-P)}{(\tilde{\xi}^{\rm CF} _{\mathbf{P}}-\xi ^{\rm B}_{\mathbf{P}-\mathbf{k}} -\tilde{\xi}^{\rm F}_{\mathbf{k}})^2}.
\end{eqnarray}

One sees clearly from Eq.~(\ref{cc}) and (\ref{dd})  that the fermionic momentum distribution function is made of two components: a Fermi distribution function of unpaired fermions and a distribution of fermions which are paired with the bosons in the molecules. The overall momentum distribution function has then a step at a momentum determined by the density of  unpaired fermions. This is the expected behavior in the molecular limit of the Bose-Fermi attraction. 
On the other hand, for a weak Bose-Fermi attraction one expects Fermi liquid theory to be valid for the Fermi component, predicting a momentum distribution function with a step at the Fermi momentum corresponding to the total fermion density. According to Luttinger's theorem,  the step remains pinned at the same momentum as for the non-interacting system, independently of the  coupling value. This is precisely what is found and discussed in Ref.~\cite{Sog13}.

Clearly, the only way to allow for such distinct behaviors  for weak and strong attraction is that a quantum-phase transition breaking down the Fermi liquid theory occurs at a certain critical coupling strength. Whether this transition coincides with the transition from the condensed phase to the normal one already studied in our previous works, or instead 
 somewhat anticipates it  within the condensed phase, is not {\em a priori} clear.   In order to answer this question, one should extend the present diagrammatic approach to the condensed phase (and/or perform extensive QMC calculations in this phase). Work along these lines is in progress.   

 Note also that at large $k$ (i.e. $k \gg k_{\rm F}$)  only the fermions belonging to the molecules contribute to the momentum distribution function. In this case 
\begin{equation}
\label{molecularwf}
n_{\rm F}(k)\to n_{\rm CF}\phi(k)^2
\end{equation}
where
\begin{equation}
n_{\rm CF}\equiv \int\!\!\!\frac{d\bf P}{(2\pi)^3} f(\tilde{\xi} ^{\rm CF} _{\mathbf{P}})
\label{ee}
\end{equation}
and, neglecting a subleading term in the expression for  $w_0$, 
\begin{equation}
\phi(k)=\sqrt{\frac{2\pi}{am_r^2}}\frac{1}{\frac{{\bf k}^2}{2m_r}+\epsilon_0}
\label{mol}
\end{equation}
is the internal wave function of the molecules (as obtained from the solution of the two-body problem).
Note further that at large $k$ also $n_{\rm B} (k)$ converges to  $n_{\rm CF} \phi (k)^2$, as it can be seen immediately from Eq.~(\ref{appequ24}).

\subsection{Chemical potentials}
The equations~(\ref{appequ24}) and (\ref{cc}) for the bosonic and fermionic momentum distributions (or for their counterparts at zero temperature) can be integrated over $k$ to obtain the boson and fermion density.
For given densities, they can be used then to get the values of the chemical potentials. In particular, it can be shown that the integration over $k$ of Eq.~(\ref{appequ24}) yields
\begin{eqnarray}
n_{\rm B}&=&\int\!\!\!\frac{d\bf k}{(2\pi)^3} n_{\rm B}(k)\\
&=&n_{\rm CF}\label{FF} + o(\epsilon_0^{-3/2}),\label{gg}
\end{eqnarray}
 where $n_{\rm CF}$ depends on the chemical potentials and temperature through Eq.~(\ref{ee}). 

Similarly, the integration over $k$ of Eq.~(\ref{cc}) yields 
\begin{eqnarray}
n_{\rm F}&=&\int\!\!\!\frac{d\bf k}{(2\pi)^3} n_{\rm F}(k)\\
&=&n_{\rm UF}\label{ff2} + n_{\rm CF} + o(\epsilon_0^{-3/2}),\label{hh}
\end{eqnarray}
where
\begin{equation}
n_{\rm UF}\equiv\int\!\!\!\frac{d\bf k}{(2\pi)^3} f(\tilde{\xi} ^{\rm F}_{\mathbf{k}}).
\end{equation}
From Eq.~(\ref{gg}) one obtains then 
\begin{equation}
\mu_{\rm CF}=\mu^0_{\rm F}(T,M^*, n_{\rm B}) + \Sigma_{\rm CF}, 
\label{mucf}
\end{equation}
where $\mu^0_{\rm F}(T,M^*, n_{\rm B})$ is the chemical potential for a free Fermi gas of temperature $T$, mass $M^*$, and density  $n_{\rm B}$, while  Eq.~(\ref{hh}) yields
\begin{equation}
\mu_{\rm F}=\mu^0_{\rm F}(T,m_{\rm F}, n_{\rm F}-n_{\rm B}) + \Sigma^0_{\rm F}, 
\label{muf}
\end{equation}
where we have used Eq.~(\ref{gg}) to replace $ n_{\rm CF}$ with  $n_{\rm B}$. The equation $\mu_{\rm B}=\mu_{\rm CF}-\mu_{\rm F}-\epsilon_0$ then yields 
\begin{eqnarray}
\mu_{\rm B}&=& \mu^0_{\rm F}(T,M^*, n_{\rm B}) - \mu^0_{\rm F}(T,m_{\rm F}, n_{\rm F}-n_{\rm B}) \nonumber\\
&+& \Sigma_{\rm CF} -\Sigma^0_{\rm F}-\epsilon_0.
\label{mub}
\end{eqnarray}

A further simplification  can be obtained by neglecting terms of order $a^2$. To this order, one can set  $M^*=M$, $\Sigma^0_{\rm F}=\frac{4\pi a}{m_r}n_{\rm B}$, and $\Sigma_{\rm CF}=\frac{4\pi a}{m_r}(n_{\rm F}-n_{\rm B})$ in the previous equations for the chemical potentials. At $T=0$ one obtains in particular:
\begin{eqnarray}
\label{equ60new}
\mu_{\rm F} &=& \frac{[6\pi^2(n_{\rm F}-n_{\rm B})]^{2/3}}{2m_{\rm F}}+\frac{4 \pi a}{m_r}n_{\rm B}, \\
\label{equ61new}
\mu_{\rm B} &=& \frac{(6 \pi ^2 n_{\rm B})^{2/3}}{2 M}+\frac{4\pi a}{m_r}(n_{\rm F}-2n_{\rm B})\nonumber \\
&-&\frac{[6 \pi ^2(n_{\rm F}-n_{\rm B})]^{2/3}}{2m_{\rm F}}-\epsilon_0 , \\
\label{equ62new}
\mu_{\rm CF}&=&\frac{(6\pi^2n_{\rm B})^{2/3}}{2M}+\frac{4 \pi a}{m_r}(n_{\rm F}-n_{\rm B}).
\end{eqnarray}
and, at this level of accuracy:
\begin{eqnarray}
k_{\rm UF}&=&\sqrt{2m_{\rm F}(\mu_{\rm F}-\Sigma_{\rm F}^0)}\nonumber\\
&=&[6\pi^2(n_{\rm F}-n_{\rm B})]^{1/3}\label{kuf}.
\end{eqnarray}
The Eqs.~(\ref{mucf})-(\ref{kuf}) show  how the T-matrix self-energy for the constituent bosons and  fermions  recovers the expected physical limit of a Fermi-Fermi mixture of dimers (molecules) and unpaired fermions mutually repelling with a scattering length $a_{\rm DF}=\gamma a$.  The T-matrix approximation yields for  the  proportionality coefficient $\gamma$ the value
$\gamma=(1+m_{\rm F}/m_{\rm B})^2/(1/2+m_{\rm F}/m_{\rm B})$, as it can be seen by 
writing the term $4\pi a/m_r$ as $2\pi a_{\rm DF}/m_{\rm DF}$, where $m_{\rm DF}=M m_{\rm F}/(m_{\rm F}+M)$ is the reduced mass of a dimer and one fermion. 
This value for $\gamma$  is only approximate and corresponds to a Born approximation for the dimer-fermion scattering.

\subsection{Comparison with the T-matrix results}\label{resultshomo}
\begin{figure}[t] 
  {\includegraphics[width=7cm]{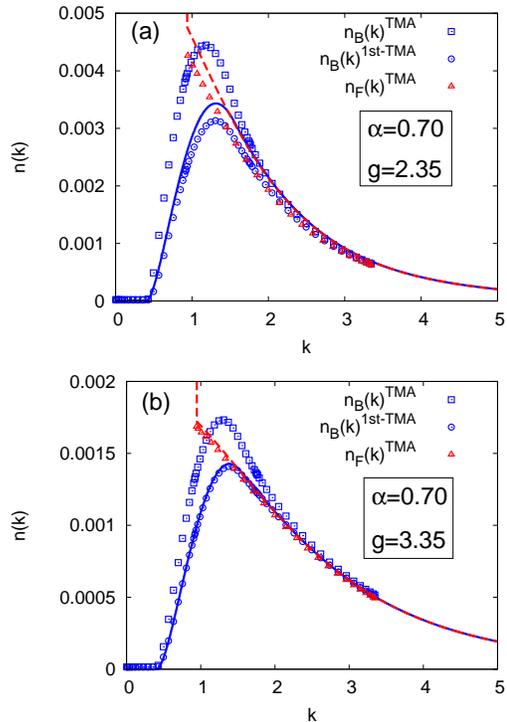}}
\caption{(Color online) Bosonic and fermionic momentum distribution function at $T=0$ for a mixture with $m_{\rm B}=m_{\rm F}$, density imbalance $\alpha$=0.70, and coupling strengths $g$=2.35 (a) and $g$=3.35 (b). The numerical results obtained by the T-matrix
self-energy (symbols) are compared with the analytic expressions in the molecular limit for $n_{\rm B}(k)$ (full curve) and $n_{\rm F}(k)$ (dashed curve) derived in the present paper.  For the bosonic distribution we present also the numerical results for the T-matrix approximation expanded to first-order in the Dyson's equation (1st-TMA) besides those obtained without expanding it (TMA). The wave-vector $k$ is in units of $k_{\rm F}$. Note that for the fermionic momentum distribution both the analytic expression and the numerical T-matrix calculation yield $n_{\rm F}(k)=1$ for $k<k_{\rm UF}$ (out of the vertical range chosen in the figure). }
\label{fig1}
\end{figure}

The asymptotic expressions for the momentum distribution function derived in Secs.~\ref{bosonic} and \ref{fermionic} can be compared with the results obtained 
by the full numerical solutions of the T-matrix set of equations. In Fig.~\ref{fig1} we present this comparison at $T=0$ for a mixture with equal masses and a density imbalance $\alpha\equiv (n_{\rm F}-n_{\rm B})/(n_{\rm F}+n_{\rm B})=0.7$ for which the  indirect Pauli exclusion effect can be seen on the bosonic momentum distribution. The two panels correspond to two different coupling values ($g=2.35,3.35$). 
For the bosonic momentum distribution, one notices that the asymptotic expression (\ref{bb})  reproduces well the T-matrix results for $k\gtrsim 2 k_{\rm F}$ as well as the presence of the empty region for $k < k_{\mu_{\rm F}}-P_{\rm CF}$, but deviates somewhat from the T-matrix results for intermediate values of $k$. This difference is due 
to the fact that the asymptotic expression  (\ref{bb}) is obtained by expanding the Dyson's equation to first-order (cf. Eq.~(\ref{appequ21})), an approximation that results to be  valid for all $k$ only for rather large values of $g$. This is confirmed by the very good agreement  with the results obtained by expanding the Dyson's equation and calculating numerically the self-energy (circles).  Clearly, for both comparisons the agreement improves when $g$ increases, albeit rather slowly for the non-expanded T-matrix. 

\begin{figure}[t] 
  {\includegraphics[width=7cm]{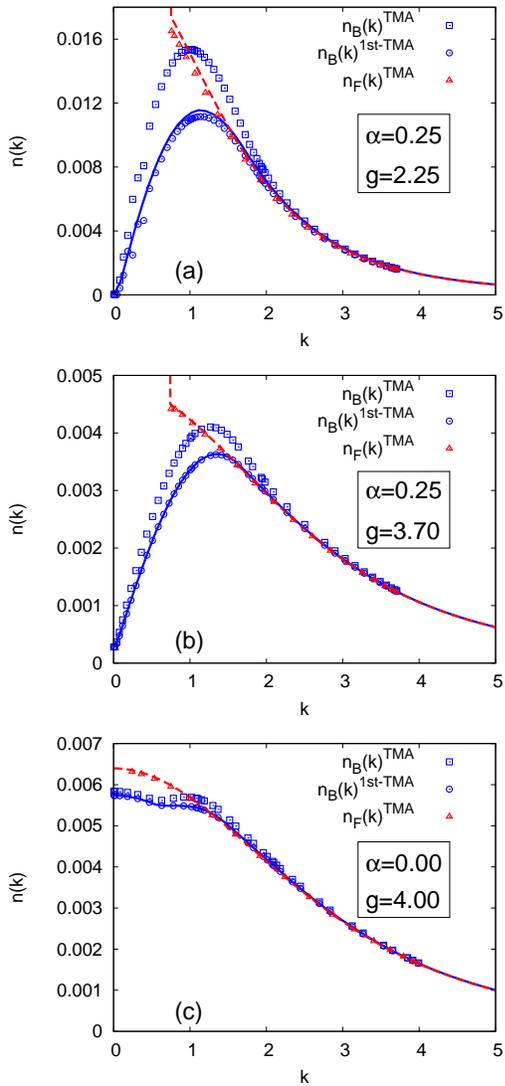}}
\caption{(Color online) Bosonic and fermionic momentum distribution function at $T=0$ for a mixture with $m_{\rm B}=m_{\rm F}$, density imbalance $\alpha$=0.25, $g$=2.25 (a)  density imbalance $\alpha$=0.25, $g$=3.70 (b), density imbalance $\alpha=0$, $g=4.0$ (c).
 The numerical results obtained by the T-matrix
self-energy (symbols) are compared with the analytic expressions in the molecular limit for $n_{\rm B}(k)$ (full curve) and $n_{\rm F}(k)$ (dashed curve) derived in the present paper.  For the bosonic distribution we present also the numerical results for the T-matrix approximation expanded to first-order in the Dyson's equation (1st-TMA) besides those obtained without expanding it (TMA). The wave-vector $k$ is in units of $k_{\rm F}$.}
\label{fig2}
\end{figure}

For the fermionic momentum distribution, instead, the asymptotic equation (\ref{dd}) compares well already with the non-expanded T-matrix (for this reason we do not present Dyson-expanded results in this case).  
 One sees that  the momentum distribution resulting from the asymptotic equation (\ref{dd}) reproduces very well the full numerical T-matrix calculation already at the lower coupling considered in Fig.~\ref{fig1}.  
Indeed, in deriving Eq.~(\ref{dd}) we have expanded the full Green's function $G_{\rm F}$ in terms of $\tilde{\Sigma_{\rm F}}\tilde{G}^0_{\rm F}$ rather than $\Sigma_{\rm F}{G}^0_{\rm F}$, a trick which, as noted above, accelerates the convergence of the Dyson's expansion in the fermionic case.

As a further check of the asymptotic expressions~(\ref{appequ21}) and (\ref{dd}) we present in Fig.~\ref{fig2} the same comparisons as  for Fig.~\ref{fig1}, but now for  two imbalances $\alpha=0.25$ and  $\alpha=0$, for which the empty region at low momenta in the bosonic momentum distribution is absent, since $P_{\rm CF} > k_{\rm UF}$.  We notice that even though also for these imbalances the asymptotic expression for the bosonic momentum distribution deviates more than the fermionic one from the T-matrix results,  the discrepancy gets smaller when the density imbalance decreases. This is due to the faster convergence of the Dyson's expansion for the bosonic Green's function at small imbalances. Indeed, one can see from Eq.(\ref{equ39a}) for the bosonic self-energy that when $P_{\rm CF}$ and $k_{ \mu_{\rm F}}$ are comparable, as it happens at small imbalances, a  partial cancellation occurs between the contributions associated to the two Fermi functions appearing in the numerator of Eq.~(\ref{equ39a}), thus making the self-energy small, and the Dyson's expansion rapidly convergent.
As a matter of fact, for the density balanced system 
at $g=4.0$ (panel (c)) one can see that the difference between the first-order expansion and the full T-matrix is indeed very small.    
 
 It is interesting to note that in this symmetric case with  $n_{\rm B}=n_{\rm F}$ (and $m_{\rm B}=m_{\rm F}$)  both the asymptotic expressions~(\ref{appequ21}) and (\ref{dd}) and the T-matrix calculations  yield slightly different occupation numbers for the boson and fermion component.
We believe that this is due to the use,  in our T-matrix approach, of bare GreenÕs functions $G^0$  multiplying the pair propagator in the expressions (\ref{selfb}) and (\ref{selff}) for the boson and fermion self-energies. In particular, if we had used a dressed fermion Green's function $G_{\rm F}$ in the place of $G^0_{\rm F}$ in the expression  (\ref{selfb}) for the boson self-energy, we  would have got a dressed fermion dispersion $\tilde{\xi}^{\rm F}_{{\bf P}-{\bf k}}$ in the place of a bare one in  Eq.~(\ref{appequ21}). Since  $k_{\rm UF}=0$ for $n_{\rm B}=n_{\rm F}$, one would have that $\tilde{\xi}^{\rm F}_{{\bf P}-{\bf k}}$  would be always positive, and the first $\Theta$ function appearing in the numerator of Eq.~(\ref{appequ21}) would be always equal to one, thus making  Eqs.~(\ref{appequ21}) and (\ref{dd}) identical for
 $n_{\rm B}=n_{\rm F}$ and $m_{\rm B}=m_{\rm F}$. The use of a bare fermionic Green's function in  Eq.~(\ref{appequ21}) subtracts instead to the integral determining the bosonic distribution function the  contribution of wave-vectors $P$ such that $|{\bf P}-{\bf k}| < k_{\mu_{\rm F}}$, with $\mu_{\rm F}= \Sigma^0_{\rm F}=\frac{4\pi a}{m_r}n_{\rm B}$ as it can be obtained from Eq.~(\ref{muf}) for $n_{\rm B}=n_{\rm F}$. This contribution vanishes in the extreme limit $g\to\infty$, but it is still finite at the value of $g$ considered in Fig.~\ref{fig2}(c), and accounts for the differences between the fermionic and bosonic distributions.  
The use of a dressed bosonic GreenÕs function in the convolution defining $\Sigma_{\rm F}$ would produce instead minor differences, due to the large and negative value of  $\mu_{\rm B}$ which makes self-energy corrections less important.

 Note finally that in the previous comparisons we used the same chemical potentials $\mu_{\rm B}$ and $\mu_{\rm F}$ calculated numerically within the T-matrix approximation  as input parameters for the asymptotic equations~(\ref{bb}) and (\ref{dd})  (the remaining parameters $M^*$, $P_{\rm CF}$ and $k_{\rm UF}$ being fully determined by the Eqs.~(\ref{equ14}),(\ref{pcf}) and (\ref{kuf})).  Alternatively, one could also use the molecular-limit expressions (47) and (48) for $\mu_{\rm B}$ and $\mu_{\rm F}$ as input parameters of the analytic calculations.  The difference in the values is small and, clearly, progressively vanishes as $g$ increases. For example, for $g=3.35$ and $\alpha=0.7$ the discrepancy amounts to  0.02 \%, 0.2 \% and 0.8 \% for $\mu_{\rm B}$, $\mu_{\rm F}$ and $\mu_{\rm CF}$, respectively.   

\subsection{Comparison between T-matrix results and Monte Carlo calculations for the bosonic momentum distribution function}\label{qmc}
In this section we present a comparison between the T-matrix results and Variational (VMC) and Fixed-node Diffusion (FN-DMC) Monte Carlo simulations obtained with a novel guiding wave function, which is a suitable symmetrization of the molecular wave function introduced in our previous work \cite{Ber13}. The details of the simulations are the same as in \cite{Ber13}, except for the trial wave function. 

Addressing the calculation of the momentum distribution of the bosons  in the molecular regime with Quantum Monte Carlo is computationally a very demanding problem, due to the need of taking care of the pairing of bosons with fermions into molecules, while simultaneously
symmetrizing with respect to the bosonic coordinates and antisymmetrizing with respect to the fermionic coordinates.
  We thus concentrate on a single choice  of the parameters, namely $g=3$, $\alpha=0.7$ and equal masses $m_{\rm F}=m_{\rm B}=m$. 
We perform our simulation with  $N_{\rm F}=40$ fermions and $N_{\rm B}=7$ bosons. These particle numbers are chosen to reduce partially finite-size effects, since the numbers of composite fermions $N_{\rm CF}=N_{\rm B}=7$ and unpaired fermions $N_{\rm UF}=N_{\rm F}-N_{\rm B}=33$ correspond to closed shells. Simulations are carried out in a cubic box of volume $L^3=N_{\rm F}/n_{\rm F}$ with periodic boundary conditions. We model the attractive interaction between bosons and fermions with a square-well potential with radius $R_{\rm BF}$ such that $n_{\rm F} R_{\rm BF}^3=10^{-7}$, and depth $V^0_{\rm BF}$ fixed by the relation $a=R_{\rm BF}(1-\tan(\kappa_{\rm BF})/\kappa_{\rm BF})$, where $\kappa_{\rm BF}=\sqrt{mV^0_{\rm BF}R^2_{\rm BF}}$. For consistency, we introduce the same repulsion between the bosons that we used in our previous work  \cite{Ber13} (even though here it would not be necessary for the stability in the molecular regime).  The repulsion is then modeled by a soft-sphere potential with radius $R_{\rm BB}=60 R_{\rm BF}$ and height $V^0_{\rm BB}$ fixed by the relation $a_{\rm BB}=R_{\rm BB}(1-\tanh(\kappa_{\rm BB})/\kappa_{\rm BB})$, where $\kappa_{\rm BB}=\sqrt{mV^0_{\rm BB}R^2_{\rm BB}}$; the Bose-Bose scattering length is set to $a_{\rm BB}=(6\pi^2 n_{\rm F})^{-1/3}$. 

In both VMC and FN-DMC the trial wave function $\Psi_T$ plays a crucial role. In VMC the sampled observables are the expectation value of quantum operators in the state defined by $\Psi_T$. In FN-DMC the amplitude of the wave function is imaginary-time evolved from $\Psi_T$, with the constraint on the nodal surface to remain pinned to the points where $\Psi_T$=0, in order to circumvent the fermionic sign problem.
We estimate the momentum distribution in VMC with $n_{\rm B}^{\rm VMC}(k)=\langle \Psi_T|\hat{n}_k|\Psi_T\rangle/\langle \Psi_T|\Psi_T\rangle$, where $\hat{n}_k$ is the number operator in momentum space averaged over momentum direction, while FN-DMC provides the mixed estimator $n_{\rm B}^{\rm DMC}(k)=\langle \Psi_T|\hat{n}_k|\Psi_0\rangle/\langle \Psi_T|\Psi_0\rangle$, where $\Psi_0$ is the long-(imaginary)-time evolution of $\Psi_T$. Both the VMC and the DMC estimates are biased by $\Psi_T$; a common way of reducing the bias is to extrapolate them via the formula $n_{\rm B}^{\rm EXT}=(n_{\rm B}^{\rm DMC})^2/n_{\rm B}^{\rm VMC}$, where the dependence on $\delta\Psi=\Psi_0-\Psi_T$ is second order, provided $\delta\Psi$ is small.

Following \cite{Ber13}, we write the guiding wave function in the molecular regime  as $\Psi_T({\bf R})=~\Phi_S({\bf R})\Phi_A({\bf R})$. Here, $\Phi_S$ is a positive Jastrow function of the particle coordinates ${\bf R}=({\bf r}_1,\dots,{\bf r}_{N_{\rm F}},{\bf r}_{1^\prime},\dots,{\bf r}_{N_{\rm B}})$ and is symmetric under exchange of identical particles. We use $\Phi_S({\bf R})=\prod_{i j^\prime}f_{\rm BF}(r_{i j^\prime})\prod_{i^\prime j^\prime}f_{\rm BB}(r_{i^\prime j^\prime})\prod_{i j}f_{\rm FF}(r_{ij})$, where the unprimed (primed) coordinates refer to fermions (bosons) and two-body spherically symmetric correlation functions of the interparticle distance are introduced. We set $f_{\rm BF}=1$, while $f_{\rm BB}$ is the solution of the two-body Bose-Bose problem with $f_{\rm BB}^\prime(L/2)=0$; $f_{\rm FF}$ is described below. Antisymmetrization in \cite{Ber13} was provided by the use of a generalized Slater determinant of the following form:
\begin{equation}
 \Phi_A^{\rm MS}({\bf R})=\left|\begin{matrix}
                        \varphi_{K_1}(1,1^\prime) & \cdots & \varphi_{K_1}(N_{\rm F},1^\prime) \\
			\vdots                   &\ddots & \vdots                  \\
			\varphi_{K_{N_{\rm M}}}(1,{N_{\rm M}}) &\cdots & \varphi_{K_{N_{\rm M}}}(N_{\rm F},{N_{\rm M}}) \\
			\psi_{k_1}(1)          &\cdots & \psi_{k_1}(N_{\rm F})     \\
			\vdots                           &\ddots & \vdots                  \\
			\psi_{k_{N_{\rm UF}}}(1)        &\cdots & \psi_{k_{N_{\rm UF}}}(N_{\rm F})
                       \end{matrix}\right|\;,
\label{eq:psiA-MS}
\end{equation}
where the molecular orbitals are defined as $\varphi_{K_\alpha}(i,i^\prime)=f_{\rm B}(|{\bf r}_i-{\bf r}_{i^\prime}|)\exp{(i {\bf K}_\alpha ({\bf r}_i+{\bf r}_{i^\prime})/2)}$, which consist of the relative-motion orbitals $f_{\rm B}$ times the molecular center-of-mass plane waves with $|K_\alpha|\le P_{\rm CF}$, and $n_{\rm CF}=P_{\rm CF}^3/6\pi^2$, while for the unpaired fermions $|k_\alpha|\le k_{\rm UF}$, with $n_{\rm UF}=k_{\rm UF}^3/6\pi^2$. The functions $f_{\rm B}$ are chosen to be the bound solutions of the two-body Bose-Fermi problem up to $\bar{R}$, matched to a functional of type $f_{\rm B}^a(r)=C_1+C_2 (e^{-\beta r}+e^{-\beta (L-r)})$ where $\bar{R}$ and $\beta$ are variational parameters and $f_{\rm B}^{a\prime}(L/2)=0$. 

The molecular orbitals appearing in the Slater determinant  \eqref{eq:psiA-MS}  are occupied by the bosons in a specific order, thus the symmetrization of the bosonic coordinates is not fulfilled. This is not a problem when calculating energies with Diffusion Monte Carlo, since the DMC pure estimator of the energy does not depend on  the trial wave function (except for the fixed nodal surface), provided there is a finite overlap of the trial wave function with the symmetric ground state.  This is the case for a finite number of particles using the non-symmetric  wave function  \eqref{eq:psiA-MS}.
 A similar approach has been successfully used in Quantum Monte Carlo studies of the equation of state of solid $^4$He \cite{Whitlock1979,cazorla2009} with the Nosanow-Jastrow wave function \cite{Nosanow1964,Hansen1968}, where the bosons are localized on specific lattice sites.
 
Bose symmetry of the trial wave function is, however, crucial when calculating the momentum distribution, which is obtained by a mixed estimator biased by the trial wave function.  A full symmetrization of the determinant \eqref{eq:psiA-MS} over all permutations of the bosons is not feasible since the number of terms to be summed scales as the factorial of $N_{\rm B}$. For this reason, we resort to an approximate strong-coupling wave function, where the symmetrization over the bosonic coordinates is performed within the molecular orbitals  appearing in a single determinant:
\begin{equation}\label{eq:psiA-MS-approx}
 \tilde{\Phi}_A^{\rm MS}({\bf R})=\left|\begin{matrix}
                        \varphi_{K_1}(1,{\bf R}_{\rm B}) & \cdots & \varphi_{K_1}(N_{\rm F},{\bf R}_{\rm B}) \\
			\vdots                   &\ddots & \vdots                  \\
			\varphi_{K_{N_{\rm M}}}(1,{\bf R}_{\rm B}) &\cdots & \varphi_{K_{N_{\rm M}}}(N_{\rm F},{\bf R}_{\rm B}) \\
			\psi_{k_1}(1)          &\cdots & \psi_{k_1}(N_{\rm F})     \\
			\vdots                           &\ddots & \vdots                  \\
			\psi_{k_{N_R}}(1)        &\cdots & \psi_{k_{N_R}}(N_{\rm F})
                       \end{matrix}\right|\;,
\end{equation}
where $\varphi_{K_\alpha}(i,{\bf R}_{\rm B})=\sum_{i^{\prime}}\varphi_{K_\alpha}(i,i^\prime)$. 

By expanding the determinant \eqref{eq:psiA-MS-approx}  it is easy to show that  $\tilde{\Phi}_A^{\rm MS}({\bf R})$  can be obtained from the original  non-symmetric wave function by summing  over all possible dispositions with repetition of the bosonic coordinates in the non-symmetric wave function \eqref{eq:psiA-MS}. The wave function $\tilde{\Phi}_A^{\rm MS}({\bf R})$  contains then all permutations of bosons, as required. It contains however also additional spurious terms where the same boson appears in many different molecular orbitals. For example, if we had $N_{\rm M}=N_{\rm B}=3$ molecular orbitals and $N_{\rm F}=5$ fermions, we would also obtain the term: $\varphi_{K_1}(1,1^\prime)\varphi_{K_2}(3,1^\prime)\varphi_{K_3}(5,1^\prime)\psi_{k_1}(2)\psi_{k_2}(4)$, where the boson $1^\prime$ is repeated. 

 These spurious terms  tend to increase the bosonic condensate, because the bosons that are not allotted to the molecular orbitals are put in a plane-wave state with zero momentum (since their spatial coordinates do not appear explicitly in these terms). In the above example, the bosons $i^\prime=2^\prime,3^\prime$ would significantly contribute to the condensate fraction, since changing their coordinates would not affect the value of that specific term.
  These terms correspond also to  the clustering of many fermions close to a single boson at a distance of order of $a_{\rm BF}$; they are then significant near resonance, where the molecular orbitals are very loose, while they are strongly suppressed in the molecular limit due to the Pauli principle, which forbids the formation of fermion clusters, thereby mitigating the unwanted effect on the bosonic condensate fraction.
 We have tried to suppress further these spurious terms, by introducing a very short-range repulsive Jastrow factor between fermions, with correlations $f_{\rm FF}$ equal to the solution of the two-body problem of a fictitious soft-sphere potential with radius $R_{\rm FF}=R_{\rm BF}$ and scattering length $a_{\rm FF}=R_{\rm FF}/10$. The above values of the parameters $R_{\rm FF}$ and $a_{\rm FF}$ were chosen so small as to avoid any significant change of the kinetic energy. It turns out that such feeble correlations do not change significantly  the momentum distribution of the bosons, either. However, they help in reducing the statistical error of the simulations; we present therefore  the results obtained by using these additional correlations, for which we have smaller error bars.
 \begin{figure} [t]
\epsfxsize=8cm
\epsfbox{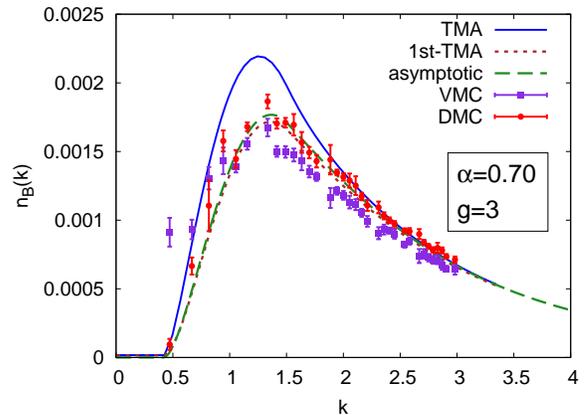}
\caption{(Color online) Comparison between T-matrix results, VMC and FN-DMC calculations for the bosonic momentum distribution at $g=3$ and $\alpha=0.7$. We also show the first-order expanded T-matrix and the asymptotic results. The wave-vector $k$ is in units of $k_{\rm F}$.}
\label{figQMC}
\end{figure}

In Fig. \ref{figQMC} we compare the VMC, DMC and T-matrix results for $n_{\rm B}(k)$ at $g=3$ and $\alpha=0.7$. Even at this value of interaction the VMC estimator gives a finite value of $n_0=n_{\rm B}(k=0)/N_{\rm B}\simeq0.06$, while the FN-DMC is able to deplete the condensate fraction down to a value compatible with zero (namely, $n_0\simeq 0.001$). One could think that obtaining a strictly zero condensate with FN-DMC and the wave function \eqref{eq:psiA-MS-approx} is in practice impossible,  because of the biased nature of the mixed estimator of the momentum distribution. The comparison between the VMC and DMC estimates hints, however, at a complete depletion of the condensate. This is  probably due to the ability of the DMC to suppress completely the energetically costly spurious terms. 

In  Fig. \ref{figQMC},  we do not report  the standard extrapolated estimator $n_{\rm B}^{\rm EXT}$ because  the presence of a spurious condensate fraction in the VMC calculation subtracts automatically weight from the rest of the distribution, thus invalidating the extrapolation procedure for all values of $k$ (including the values of $k$ where the VMC and FN-DMC are close to each other, for which the extrapolation procedure could appear justified).  
The DMC calculation confirms the suppression of the bosonic momentum distribution  at low $k$, in particular the DMC results appear to  follow the T-matrix curve from $k\simeq 1$ down to the value of $k$ where the momentum distribution is predicted to vanish according to the T-matrix calculation.
The DMC calculation agrees well with the  T-matrix results also at high momenta ($k \gtrsim 2$).
Some deviations occur in the intermediate region  $1< k < 2$, where the DMC seem closer to the first-order expanded T-matrix curve rather than the full T-matrix curve.  We regard this better agreement with the expanded T-matrix at intermediate $k$ as fortuitous. On the one hand,  an extrapolation of the VMC and DMC results would increase the values of the momentum distribution in this region, making it closer to the T-matrix curve.  On the other hand, the relative motion molecular orbital $f_{\rm B}$ strongly affects the nodal surface and thus the momentum distribution. It can be argued that refining its parametrization would modify the occupation of intermediate momenta.  Addressing quantitatively these issues and reducing the error bars, especially for $k<k_{\rm F}$, would require,  however, an extremely large computational effort.

\section{Trapped system}
\label{trapped}
The equations derived in the previous section for the momentum distribution functions and for the chemical potentials (and derived quantities, such as $P_{\rm CF}$ and $k_{\rm UF}$) can be used to describe also a Bose-Fermi mixture trapped in an external potential  whenever the particle number is sufficiently large to make a local density approximation accurate.  For the particle numbers of order $10^5$-$10^7$  typically used in experiments with ultracold trapped gases this condition is fully satisfied.
The effect  of the trapping potential is then taken into account by replacing the chemical potentials $\mu_{\rm B,F} \to \mu_{\rm B,F} - V_{\rm B,F}(r)$ wherever they appear in the expressions derived in the previous section for homogeneous gases. Here, $V_{\rm B,F}(r)=\frac{1}{2}\omega_{\rm B, F} r^2$ is the harmonic trapping potential acting on the boson and fermion species, respectively (for definiteness we assume the same trap frequency $\omega$ for both species). 
The local quantities derived in this way can be integrated over $r$ to obtain the corresponding trap-averaged quantities.

\begin{figure}[t]
\epsfxsize=8cm
\epsfbox{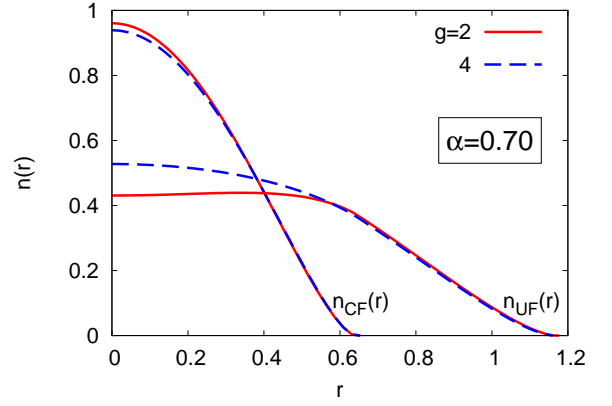}
\caption{(Color online) Density profiles of composite fermions (CF) and unpaired fermions (UF) for a mixture with equal masses and population imbalance $\alpha=0.7$ for  coupling values $g$=2.0, 4.0. Density is in units of $N_{\rm F} R_{\rm F}^3$, while $r$ is in units of the Fermi radius $R_{\rm F}\equiv[2E_{\rm F}/(m_{\rm F}\omega^2)]^{1/2}$.}
\label{fig4}
\end{figure}

We will be interested in particular in the calculation of the trap-averaged momentum-distribution function $n_{\rm B}^{\rm trap}({\bf k})$, with the aim of determining the best 
conditions for the observation of the ``indirect Pauli exclusion effect"  in trapped gases.
The local bosonic momentum distribution function in the molecular limit is then given by
\begin{equation}
n_{\rm B}({\bf k},r)=w_0\! \int\!\!\frac{d\bf P}{(2\pi)^3}
\frac{\Theta(\xi ^{\rm F}_{\mathbf{P}-{\bf k}}(r)) \Theta(P_{\rm CF}(r)-P)}
{[\xi ^{\rm B}_{\bf k}(r)+\xi ^{\rm F}_{{\bf P}-{\bf k}}(r)-\tilde{\xi}^{\rm CF} _{\bf P}(r)]^2}
\label{nbktrap}
 \end{equation}
 where $\xi ^{\rm B,F}_{\bf k}(r)=\xi ^{\rm B,F}_{\bf k}+V_{\rm B,F}(r)$,  while $\tilde{\xi}^{\rm CF} _{\bf P}(r)=\frac{P^2}{2M}-\mu_{\rm CF} +V_{\rm CF}(r) +\Sigma_{\rm CF}(r)$, with $\Sigma_{\rm CF}(r)=\frac{2\pi a_{\rm DF}}{m_{\rm DF}}n_{\rm UF}(r)$, $V_{\rm CF}(r)=V_{\rm B}(r)+V_{\rm F}(r)$ and we have defined the density of unpaired fermions $n_{\rm UF}(r)=n_{\rm F}(r)-n_{\rm B}(r)$.

 The trap-averaged quantity is then readily obtained  by integrating over $r$:
 \begin{equation}
 n_{\rm B}^{\rm trap}({\bf k})=\int d^3r \; n_{\rm B}({\bf k},r).
 \label{trapav}
 \end{equation}
 
 The chemical potentials $\mu_{\rm B,F}$ (and thus $\mu_{\rm  CF}=\mu_{\rm B}+ \mu_{\rm F}+\epsilon_0$) appearing in Eq.~(\ref{nbktrap}) need to be determined by the number equation, obtained by integrating over $r$ the corresponding densities $n_{\rm B,F}(r)$. Since in the molecular limit  all bosons are inside the molecules,  it is  physically more transparent to work in terms of the molecular and unpaired fermion densities,  $n_{\rm CF}(r)=n_{\rm B}(r)$ and $n_{\rm UF}(r)$, respectively.
 
 From the Eqs.~(\ref{equ60new}-\ref{equ62new}) one gets
 \begin{eqnarray}\label{equ71}
n_{\rm CF}(r)&=&\frac{1}{6\pi ^2}\lbrace2M[\mu_{\rm CF}-V_{\rm CF}(r)
-\frac{2\pi a_{\rm DF}}{m_{\rm DF}} n_{\rm UF}(r)]\rbrace^{3/2}\nonumber\\
&&\\
n_{\rm UF}(r)&=&\frac{1}{6\pi ^2}\lbrace 2m_{\rm F}[\mu _{\rm F}-V_{\rm F}(r)-\frac{2\pi a_{\rm DF}}{m_{\rm DF}}n_{\rm CF}(r)]\rbrace^{3/2}, \nonumber \\
&&\label{equ72}
\end{eqnarray}
from which the chemical potentials $\mu_{\rm F}$ and $\mu_{\rm CF}$ are obtained by fixing the total number of composite fermions $N_{\rm CF}=N_{\rm B}$ and
 unpaired fermions $N_{\rm UF}=N_{\rm F}- N_{\rm B}$.

 \begin{figure}[t]
\epsfxsize=8cm
\epsfbox{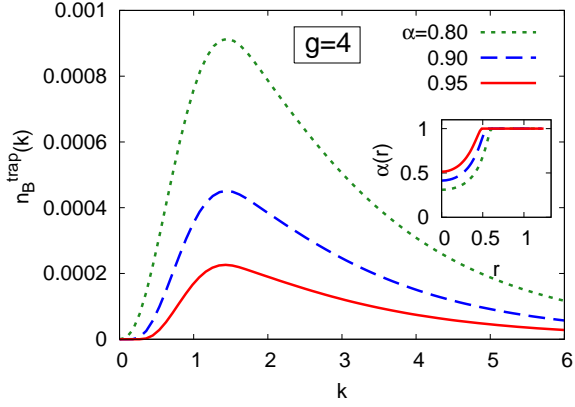}
\caption{(Color online) Trap-averaged bosonic momentum distribution function $n_{\rm B}^{\rm trap}(k)$ for $g=4$, $m_{\rm B}=m_{\rm F}$, and different values of the population imbalance. Inset: local density imbalance $\alpha(r)$.  The wave-vector $k$ is in units of $k_{\rm F}$, while $r$ is in units of $R_{\rm F}$. }
\label{fig5}
\end{figure}

Figure {\ref{fig4} reports as an example the density profiles for a mixture with equal masses and population imbalance $\alpha\equiv(N_{\rm F}-N_{\rm B})/(N_{\rm F}+N_{\rm B})=0.7$ for two coupling values $g$=2.0, 4.0. Here, as for the homogeneous case, we have defined $g=(k_{\rm F} a)^{-1}$ and  $k_{\rm F}=(2m_{\rm F}E_{\rm F})^{1/2}$, but with   $E_{\rm F}=(6N_{\rm F})^{1/3}\omega$ in the trapped case.  Note that here we are using the exact relation between $a_{\rm DF}$ and $a$, as obtained from the solution of the three-body problem~\cite{iskin}. The behavior of the density profiles is consistent with analogous plots reported previously for Fermi-Fermi mixtures  (albeit with equal populations~\cite{Roth01,Sogo02}). 

\begin{figure}[t]
\epsfxsize=8cm
\epsfbox{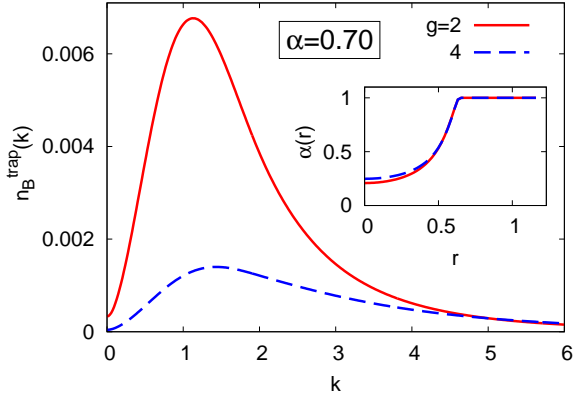}
\caption{(Color online) Trap-averaged bosonic momentum distribution function $n_{\rm B}^{\rm trap}(k)$ for $\alpha$=0.7, $m_{\rm B}=m_{\rm F}$, and coupling strength $g=2,4$. Inset: local density imbalance $\alpha(r)$.  
The wave-vector $k$ is in units of $k_{\rm F}$, while $r$ is in units of $R_{\rm F}$.}
\label{fig6}
\end{figure}

\begin{figure}[t]
\epsfxsize=7.5cm
\epsfbox{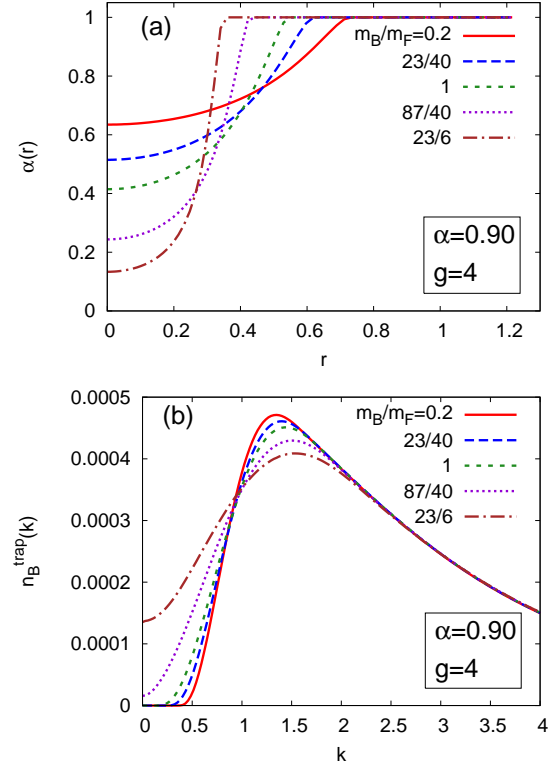}
\caption{(Color online) (a) Local density imbalance $\alpha(r)$, and (b) trap-averaged bosonic momentum distribution function $n_{\rm B}^{\rm trap}({\bf k})$ at $g=4$ and $\alpha=0.9$
 for different values of the mass ratio $m_{\rm B}/m_{\rm F}$. The wave-vector $k$ is in units of $k_{\rm F}$, while $r$ is in units of $R_{\rm F}$.}
\label{fig7}
\end{figure}

Once the chemical potentials are obtained by inverting the number equations using the above density profiles, the trap-averaged momentum distribution function is calculated with Eqs.~(\ref{nbktrap}) and (\ref{trapav}). Figure~\ref{fig5} reports the trap-averaged bosonic momentum distribution function $n_{\rm B}(k)$ for three different population imbalances at $g=4$ and equal masses. One observes that the depleted region at low momenta is visible also for the trapped system, provided the overall population imbalance is sufficiently high. In order to interpret these results, we note first that the previous equations for a homogeneous mixture imply that the depleted region at low momenta appears when the radius of the Fermi sphere of the unpaired fermions exceeds that of the composite fermions.  In the molecular limit this occurs when $n_{\rm B}< n_{\rm F}/2$ or, equivalently,  for a density imbalance $\alpha > 1/3$.  It follows then that in order to have the empty region at low momenta also in the trapped case, the local density imbalance $\alpha(r) \equiv (n_{\rm F}(r) - n_{\rm B}(r))/(n_{\rm F}(r) + n_{\rm B}(r))$ should be larger than $1/3$ all over the trap. One sees in the inset of Fig.~\ref{fig5}  that this condition is indeed verified for the three cases considered there.   
It is clear then that, in order to maximize the indirect Pauli-exclusion effect on the bosonic momentum distribution, one has to get large values of $\alpha(r)$ across the trap.
Quite generally,  the local density imbalance depends on three different physical parameters: the global population imbalance $\alpha$, the boson-fermion coupling $g$ and the mass ratio $m_{\rm B}/m_{\rm F}$. One has then to tune appropriately these parameters.  Obviously, a large global population imbalance increases the local one,  as it is also evident  from Fig.~\ref{fig5}.
 
Figure~\ref{fig6} shows instead that, for a given population imbalance,  increasing the coupling strength  $g$ has a modest effect on $\alpha(r)$, while the momentum distribution function is reduced at low $k$ (and increased at large $k$, outside the range shown in Fig.~\ref{fig6}), reflecting the behavior of the internal molecular wave function (\ref{mol}).

The dependence on the mass ratio $m_{\rm B}/m_{\rm F}$ is studied finally in Fig.~\ref{fig7}. One can see that for given coupling strength and population imbalance (here $g=4$ and $\alpha=0.9$), decreasing the mass ratio $m_{\rm B}/m_{\rm F}$ increases the local population imbalance, thus making more evident the 
presence of the empty region at low momenta.  Note that three out of the five mass ratios considered in Fig.~\ref{fig7} correspond to the mixtures  $^{23}$Na-$^{40}$K,
$^{87}$Rb-$^{40}$K, and $^{23}$Na-$^{6}$Li,  of relevance to current experiments~\cite{Wu12}, \cite{Bloom13}, \cite{Heo12}.  It should be stressed, in this respect, that while the Feshbach resonances used for the first two mixtures are broad~\cite{Wu12,Bloom13}, the one currently used for the $^{23}$Na-$^{6}$Li  mixture is narrow~\cite{Heo12}.  Therefore, while for the first two mixtures the single-channel description adopted in the present work is fully adequate~\cite{Sim05},  for the last one  our analysis has to be regarded as  more qualitative.
We notice finally that out of these three mixtures,   the  $^{23}$Na-$^{40}$K mixture looks as as the most  promising one for the experimental observation of the indirect Pauli exclusion effect, since it leads to a wider depleted region in the bosonic momentum distribution.

\section{Concluding Remarks}\label{conclusions}
In summary, we  have shown how, within a T-matrix diagrammatic approach, a Fermi-Fermi mixture emerges effectively from a Bose-Fermi mixture for sufficiently strong attraction.
In this limit, we have derived  simple expressions for the bosonic and fermionic self-energies, momentum distribution functions, and chemical potentials. 
In particular, we have obtained an expression for the bosonic momentum distribution function that shows  analytically the presence of a completely depleted region at low momenta when $n_{\rm B} < n_{\rm F}/2$.  The occurrence of this region is the fingerprint of what we called  the {\em indirect Pauli exclusion effect}.
We have confirmed the presence of such a region also with a dedicated QMC simulation. This required us to address the non-trivial problem of symmetrizing with respect to the bosonic coordinates a trial wave function where the bosons are correlated with fermions in a molecular bound state. To this end, we have introduced a wave function where the symmetrization is performed within each molecular orbital rather than globally, as to keep the computing time manageable. 

Finally, we have discussed the possibility of observing the indirect Pauli exclusion effect in a trapped system, by extending within a local density approximation our calculations to such an inhomogeneous situation. We have found that mixtures where the bosons are lighter than the fermions  enhance the size and visibility of the depleted region:  the mixture  $^{23}$Na-$^{40}$K currently under study at the MIT~\cite{Wu12} appears particularly interesting in this respect.

\acknowledgments
G.B. acknowledges useful discussions with D.E. Galli. Part of the QMC simulations were performed on the SuperB cluster at EPFL.


\begin{thebibliography}{99}
\bibitem{footnote1}  At zero temperature and for $P < k_{\mu_{\rm F}} M/m_{\rm F} $,  the branch-cut is shifted to ${\rm Re} \, z \ge \xi^{\rm B}_{P-k_{\mu_{\rm F}}}$, where $k_{\mu_{\rm F}}=\sqrt{2 m_{\rm F} \mu_{\rm F}}$. It is easy to verify that  $\xi^{\rm B}_{P-k_{\mu_{\rm F}}}\ge 2\mu + P^2/(2M)$.
\end{thebibliography}
\end{document}